\documentclass[review]{elsarticle}

\usepackage{lineno,hyperref}
\usepackage{lgreek}
\usepackage{amsfonts}
\usepackage{amsmath}
\usepackage{amssymb}
\usepackage[final]{pdfpages}

\begin{document}

\begin{frontmatter}

\title{Solvent-induced morphological transitions in methacrylate-based block-copolymer aggregates}

\author{Gerardo Campos-Villalobos}
\author{Flor R. Siperstein}
\author{Arvin Charles}
\author{Alessandro Patti\corref{mycorrespondingauthor}}
\cortext[mycorrespondingauthor]{Corresponding author}
\ead{Alessandro.Patti@manchester.ac.uk}
\address{Department of Chemical Engineering and Analytical Science, University of Manchester, Sackville Street,
Manchester M13 9PL, U.K.}




\begin{abstract}
Poly(ethylene oxide)-\textit{b}-poly(butylmethacrylate) (PEO-\textit{b}-PBMA) copolymers have recently been identified as excellent building blocks for the synthesis of hierarchical nanoporous materials. Nevertheless, while experiments have unveiled their potential to form bicontinuous phases and vesicles, a general picture of their phase and aggregation behavior  is still missing. By performing Molecular Dynamics simulations, we here apply our recent coarse-grained model of PEO-\textit{b}-PBMA to investigate its self-assembly in water and tetrahydrofuran (THF) and unveil the occurrence of a wide spectrum of mesophases. In particular,  we find that the morphological phase diagram of this ternary system incorporates bicontinuous and lamellar phases at high copolymer concentrations, and finite-size aggregates, such as dispersed sheets or disk-like aggregates, spherical vesicles and rod-like vesicles, at low copolymer concentrations. The morphology of these mesophases can be controlled by tuning the THF/water relative content, which has a striking effect on the kinetics of self-assembly as well as on the resulting equilibrium structures. Our results disclose the fascinating potential of PEO-\textit{b}-PBMA copolymers for the templated synthesis of nanostructured materials and offer a guideline to fine-tune their properties by accurately selecting the THF/water ratio.
\end{abstract}


\end{frontmatter}


\begin{sloppypar}

\section{Introduction}

In the early 1900's, the pioneering experiments by McBain~\cite{mcbain1920} and Hartley~\cite{hartley1938} disclosed a realm of intriguing morphologies resulting from the spontaneous self-assembly, in selective solvents, of a particular family of molecules. These molecules, consisting of a solvophilic and a solvophobic domain, are generally referred to as amphiphiles, from the Greek \begin{greek} >'amfis\end{greek} (both) and \begin{greek} fil'ia \end{greek} (love). The technological implications of these observations have become clearer and clearer over the last few decades, with the development of synthesis techniques transforming relatively simple molecular building blocks into structured supramolecular aggregates and these into \textit{ad hoc} templates for highly ordered nanoporous materials~\cite{beck1992}. Unveiling the physico-chemical principles underpinning the microstructural transitions that occur in micellar solutions, dating back to Debye~\cite{debye1951}, has been crucial to identify the key factors controlling the formation of equilibrium aggregates: amphiphile's architecture and concentration, pH, temperature, solvent and additives ~\cite{cates1990}. The coordinated action of these factors can be concisely summarized by the so-called packing parameter, $p=\nu/al$, where $\nu$ and $l$ are the volume and effective length of the solvophobic block, respectively, while $a$ is the area per solvophilic head group. In particular, lamellae and vesicles are expected at $1/2 < p \le 1$, whereas rod-like and spherical micelles at $1/3<p  \le 1/2$ and $p \le 1/3$, respectively ~\cite{israelachvili2015}. 

In particular, when diblock-copolymers are added to a solvent that has a selective affinity for one of the blocks, the self-assembly of the other block sparks the formation of colloidal clusters or micelles. In the most general case, these aggregates consist of a solvophobic core (hydrophobic, in aqueous solutions) and a solvophilic (hydrophilic) corona, but more complex copolymer's architectures, comprising three or more blocks, can form aggregates with more than just two separate domains. If the volume occupied by the corona is significantly larger than that occupied by the core, then the resulting aggregates are generally referred to as "star-like" micelles. By contrast, aggregates with a bulky core and a thin corona are defined as "crew-cut" micelles ~\cite{halperin1992}. Today, more than 20 stable morphologies obtained via self-assembly of block-copolymers (BCPs) have been reported~\cite{mai2012}, approximately three times more than those firstly identified almost 25 years ago by Zhang and Eisenberg~\cite{zhang1995}. Some morphologies, such as spherical or rod-like micelles, are more frequently observed, while others - hexagonally packed hollow hoops~\cite{zhang1997}, helical micelles~\cite{zhong2008}, disk-like micelles (also called "hamburger micelles")~\cite{li2006} and multilamellar vesicles~\cite{shen2000} - are more exotic. In general, the BCP assemblies exhibit a higher stability and longevity compared to those formed by small surfactants and consequently have attracted considerable attention as drug-delivery systems in biomedicine~\cite{rosler2012,buwalda2018, yang2010}, templates for fabricating capacitors with increased charge storage capacity in microelectronics~\cite{black2001, kim2009} and nanoreactors and multiple stimuli-responsive biomaterials~\cite{ discher2002, palivan2016}.

The self-assembly of BCPs in solution is dictated by principles analogue to those valid for small surfactants. Nevertheless, extra elements of complexity are present, such as the long-time relaxation dynamics and higher degree of hydrophobicity of polymeric chains~\cite{won2003, jain2003}. Consequently, the synthesis method to obtain BCP's mesophases tightly depends on the composition of the copolymer being employed as well as on the size of its two blocks. Usually, aggregates of BCPs with relatively bulky hydrophobic blocks, such as polystyrene-\textit{b}-poly(acrylic acid), which show an especially rich variety of morphologies~\cite{zhang1995}, are prepared via the so-called \emph{solvent switch} or co-solvent method~\cite{zhang1996}. This method consists in dissolving a small amount of BCP in an organic solvent, such as dioxane, \textit{N,N'}- dimethylformamide (DMF) or tetrahydrofuran (THF), that is a good solvent for both blocks. Subsequently, water, a selective solvent for the hydrophilic block, is slowly added to this solution up to a content (usually between 25 and 50 wt\%) that is significantly larger than the water content at which micellization starts~\citep{mai2012}. Finally, the so-obtained aggregates are quenched in excess of water in order to freeze all the kinetic processes and morphologies. Eisenberg and coworkers found that various morphologies could be obtained for a given copolymer architecture and concentration by tuning solely the common/selective solvent ratio~\cite{shen1999}, thus offering an alternative path to control the properties of the nanostructures at equilibrium~\cite{choucair2003}. In order to exploit the beneficial implications of such an additional degree of freedom, it becomes essential to understand how altering the medium quality determines the kinetics and thermodynamics of BCP self-assembly. To this end, one needs to gain a full insight into the nature of the molecular interactions, which is all but trivial and nevertheless of fundamental importance.

Statistical-mechanical theories and computer simulation, including molecular Dynamics (MD), Dissipative Particle Dynamics (DPD) and Monte Carlo (MC) methods, have proven to be especially effective in addressing this challenging task and offered a clearer picture of the physico-chemical effects driving the phase and aggregation behavior of BCPs. Over the last two decades, most of the computational works have predominantly focused on the self-assembly of low molecular weight surfactants, lipids and BCPs in water solutions, and on the estimation of their critical micelle concentration (CMC), aggregation number, aggregate size and solution rheology~\cite{larson1989, goetz1998, siperstein2003, srinivas2004, padding2005, patti2007, patti2007phase, jorge2008, patti2009_1, patti2009_2, patti2010, lee2013, wang2017st}. More recently, mean field theories have been applied to investigate the kinetics of molecular exchange in micellar solutions~\cite{garcia2017_1, garcia2017_2}. By contrast, only little attention has been paid to the self-assembly of amphiphiles in solutions of two solvents. In particular, Li and coworkers have recently applied the simulated annealing MC method to investigate the phase behavior of generic amphiphilic AB diblock copolymers in mixtures of selective and common solvents~\cite{wang2017}. The authors observed that by increasing the amount of selective solvent, self-assembled structures experience a number of morphological transformations that follow a sphere$\rightarrow$rod$\rightarrow$ring/cage$\rightarrow$vesicle sequence. 

Most of these studies have been performed by employing oversimplified potential models that neglect important chemical details of amphiphilic BCPs. Molecular simulations of fully atomistic models remain prohibitively expensive because the timescale for spontaneous self-assembly is usually too long (microseconds) and the required system size too large (several hundreds of thousands of atoms) to be systematically investigated using the currently available computational power. An alternative route is based on high-level coarse-grained (CG) potential models that are built via a multiscale simulation approach. More specifically, fully atomistic models of smaller systems are employed to estimate some well-selected properties that are used to identify the most suitable parameters of the CG model's force field~\cite{srinivas2004, giunta2019, lee2012}. In this work, we employ our recently developed CG model for methacrylate-based copolymers~\cite{campos2019} to investigate the self-assembly of low-molecular weight poly(ethylene oxide)-\textit{b}-poly(butylmethacrylate) (PEO-\textit{b}-PBMA) copolymers in mixtures of water and THF. This ternary system has been recently reported to exhibit a wide variety of self-assembled nanostructures, including bicontinuous polymer nanospheres, dispersed sheets, vesicles and octopus-like assemblies~\cite{mckenzie2013, mckenzie2015}. By MD simulations, we map their phase diagram by mimicking the solvent-switch method and observe a wide spectrum of interesting mesophases at different common/selective solvent ratios. We also discuss the mechanisms driving the kinetics of morphological transitions that are determined by changes in the solvent composition. Our findings highlight the importance of solvent correlations on the intra-aggregate chain structure and resulting equilibrium morphology.

\section{Methods}

\subsection{Model}
\begin{figure}
\begin{center}
\includegraphics[scale=0.29]{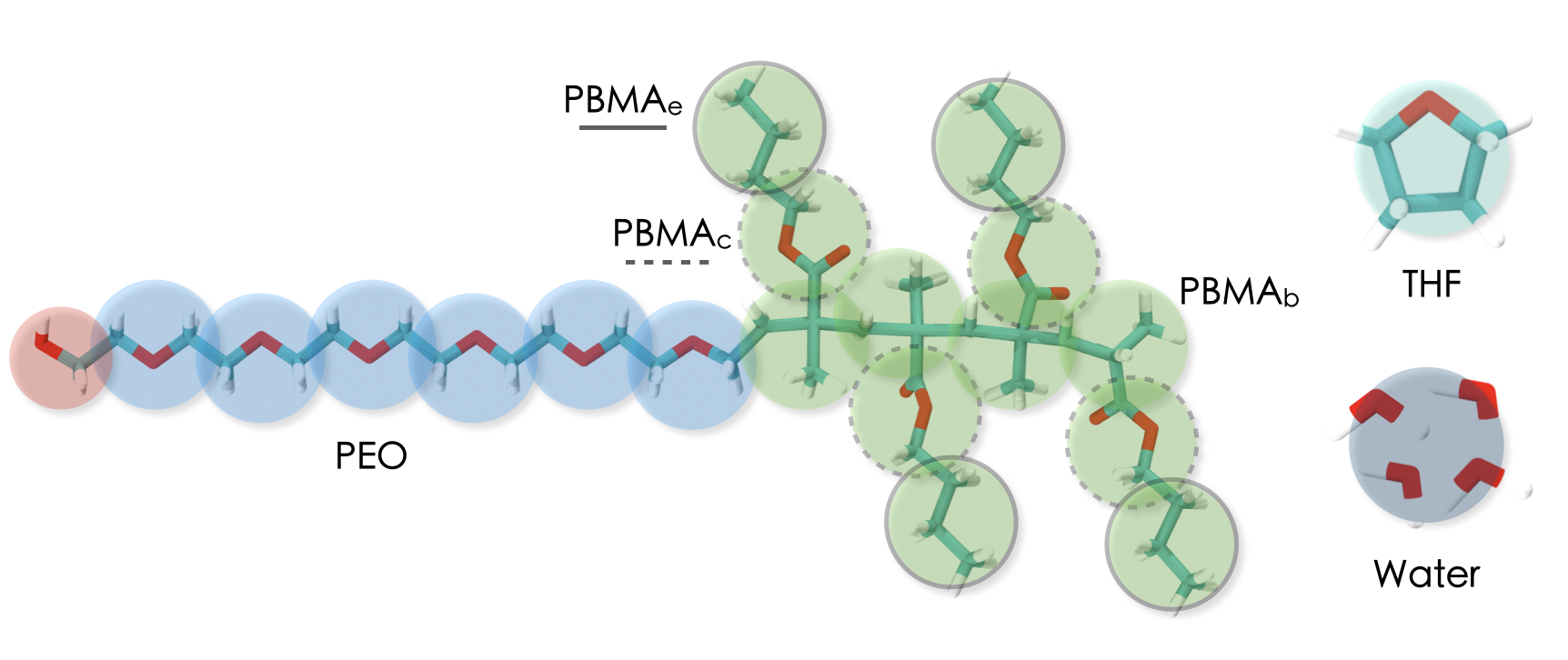}
\caption{Coarse-grained and underlying atomistic representations of PEO$_{6}$-\textit{b}-PBMA$_{4}$, water and THF.  At the CG level, a PEO segment of $n$ monomers consist of $n+1$ sites due to the terminal HO-CH$_{2}$- group. The PBMA$_{\textrm{e}}$, PBMA$_{\textrm{c}}$ and PBMA$_{\textrm{b}}$ sites denote terminal, bridging, and backbone beads of the PBMA repeating unit, respectively. THF is represented by a single bead and the standard MARTINI 4-to-1 water model is employed. Red, white and light blue solid segments in the atomistic model represent oxygen, hydrogen and carbon atoms, respectively. The interested reader is referred to Ref.~\cite{campos2019} for additional details on the model.}
\label{mappings}
\end{center}
\end{figure}

Classical MD simulations were performed using CG-scale representations of PEO-\textit{b}-PBMA copolymers, water and THF. Two different methacrylate-based BCPs, PEO$_{6}$-\textit{b}-PBMA$_{4}$ and PEO$_{12}$-\textit{b}-PBMA$_{10}$, have been modelled in this work. Their interactions were described via inter- and intra-molecular potentials of the MARTINI force-field family~\cite{marrink2007}. In particular, we adopted our recently refined parameters, which allow for the reproduction of the structural and thermodynamic properties of the BCPs in melt and solutions~\cite{campos2019}. In this scheme, the PBMA monomer is represented by three CG beads, whereas the PEO monomer and THF are modelled by a single site, as reported in Fig.~\ref{mappings} for PEO$_{6}$-\textit{b}-PBMA$_{4}$.

\noindent Pairs of non-bonded CG beads interact with each other via a truncated and shifted Lennard-Jones (LJ) potential:

\begin{equation}
U^{\textrm{LJ}}(r) = 4 \epsilon \left[  \left(\frac{\sigma}{\epsilon} \right)^{12} - \left(\frac{\sigma}{\epsilon} \right)^{6} \right]-U(r_c),
\end{equation}

\noindent where $\sigma$ and $\epsilon$ are, respectively, the length- and energy-scale parameters of the pair interaction, $r$ is the separation distance between pairs of spherically-symmetric CG sites, and $U(r_c)$ is the value of the LJ potential at the cut-off radius, $r_c$. Within the BCP chains, intramolecular interactions acting on the centres of bonded sites are described using a harmonic bond-stretching potential:

\begin{equation}
U^{\textrm{bond}}(l) = \frac{1}{2} K_{l} \left( l - l_{0} \right)^{2}
\end{equation}

\noindent with $K_{l}$ the bond force constant, and $l$ and $l_{0}$ the instantaneous and equilibrium bond distances, respectively. Similarly, the angle-bending between triplets of connected beads is modelled via a harmonic potential:

\begin{equation}
U^{\textrm{angle}}(\theta) = \frac{1}{2} K_{\theta} \left(  \cos \theta - \cos \theta_{0} \right)^{2}
\end{equation}

\noindent where $K_{\theta}$ is the angle  force constant, and $\theta$ and $\theta_{0}$ the instantaneous and equilibrium angle-bending values, respectively. The list of all the parameters employed in this work is available to the interested reader in the Supporting Information.

\subsection{Molecular Simulations}
We have investigated the aggregation behavior of  PEO$_{6}$-\textit{b}-PBMA$_{4}$ and PEO$_{12}$-\textit{b}-PBMA$_{10}$ BCPs in mixtures of water and THF of varying composition. The PEO content of these BCPs, defined as the ratio of the mass of the PEO block and the whole BCP chain,  $M_{\textrm{PEO}}/M_{\textrm{BCP}}$, are 0.31 and 0.27, respectively. These BCPs have been selected to study mainly the effect of chain length, and to a lesser extent the hydrophobic/hydrophilic ratio on the equilibrium morphologies. In order to accurately map the phase diagrams of such ternary systems, we performed simulations at $0.05 \leq \omega_{\textrm{BCP}} \leq 0.50$, where  $\omega_{\textrm{BCP}}$ is the copolymer mass fraction. At each value of  $\omega_{\textrm{BCP}}$, different solvent ratios, $f_{\textrm{w}}\equiv \omega_{\textrm{w}}/\left(\omega_{\textrm{w}} + \omega_{\textrm{THF}}\right)$, with $\omega_{\textrm{w}}$ and $ \omega_{\textrm{THF}}$ the water and THF mass fractions, respectively, have been analyzed. It follows from the above definitions that $\omega_{\textrm{w}} + \omega_{\textrm{THF}} +  \omega_{\textrm{BCP}} = 1$. Each state, characterized by fixed values of $\omega_{\textrm{BCP}}$ and  $f_{\textrm{w}}$, was simulated independently without altering the compositions from the original (initial) configuration. Experimentally, the composition of the system changes by stepwise addition of water~\cite{mai2012}. Therefore, our simulations do not mimic the full synthesis process, but they are expected to be consistent with intermediate steps in a solvent-switch experiment being performed at slow rates of water addition, where the systems at each intermediate composition are allowed to reach equilibrium.

Initial configurations were obtained by random spatial distribution and orientation of chain and solvent molecules in cubic simulation boxes with periodic boundaries using PACKMOL~\cite{martinez2009}. The number of BCP chains was set to $N_{\textrm{BCP}}= 750$ and the amount of solvent was estimated so as to obtain the desired mixture composition. The packing of molecules in the initial state was performed at relatively low density ($\sim 800$ kg m$^{-3}$) in order to avoid molecular overlaps. Energy minimization was applied prior to equilibration and production runs to eliminate unrealistic high-energy structures using the steepest-descent method. MD simulations were carried out in the isothermal-isobaric (NPT) ensemble using the GROMACS 5.0.4 package~\cite{van2005}. The temperature, $T=300$ K, was controlled by means of the stochastic velocity-rescale algorithm by Bussi \textit{et al.}~\cite{bussi2009} with a coupling constant of 1.0 ps. The Berendsen barostat~\cite{berendsen1984} was employed to restrict pressure fluctuations about the equilibrium value of $P=1$ bar with a relaxation constant of 3.0 ps and a standard compressibility of 5$\times 10^{-5}$ bar$^{-1}$. With this algorithm, the initial configurations with densities around 800 kg m$^{-3}$ were rapidly compressed to their equilibrium values during the first hundreds of picoseconds. Integration of the classical equations of motion was accomplished by means of the leapfrog algorithm with a timestep of 20 fs. Dispersion interactions were truncated at $r_{\textrm{c}}=1.2$ nm as in the standard MARTINI force-field. Long-range corrections, compensating the potential truncation, were added to the total energy and pressure. In order to guarantee equilibration and discard the occurrence of artificial non-ergodic structures, all simulations were run for 7 $\mu$s and for each state we performed two independent simulations starting from different velocity distributions and molecular coordinates. When necessary, reversibility along the BCP iso-concentration lines (direct transition) was tested by extra MD runs using the final configuration of a state point as the initial one for a neighbouring state and manually modifying the solvent composition. All the morphologies reported in the remaining of this article were stable for at least 3 $\mu$s.

\section{Results}
\begin{figure}
\begin{center}
\includegraphics[scale=0.27]{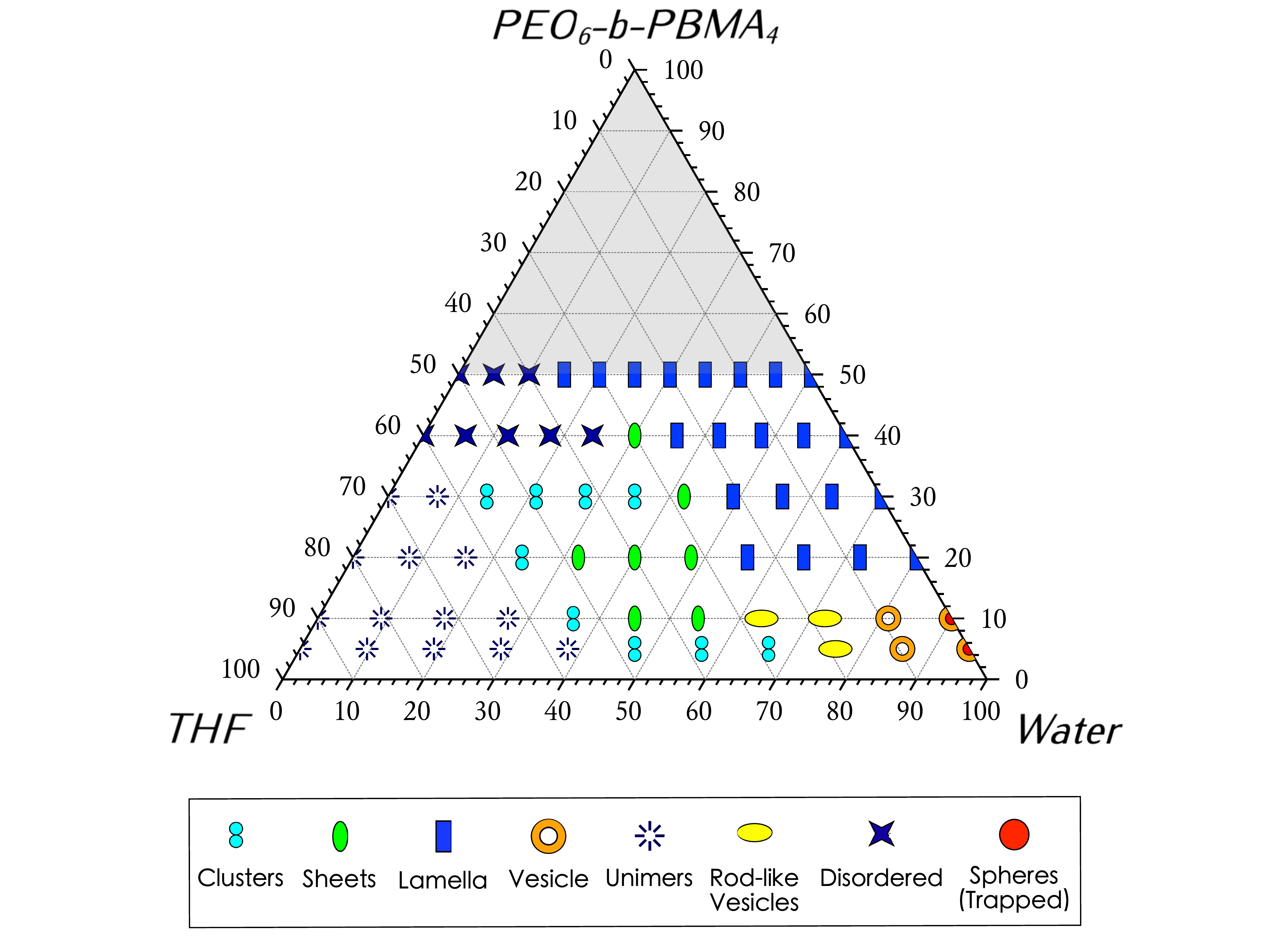}
\caption{Morphological diagram of PEO$_{6}$-\textit{b}-PBMA$_{4}$ in water and THF at $T=300$ K. The compositions are given in wt$\%$. The symbols refer to different self-assembled structures as indicated in the legend panel. The shaded area has not been studied.}
\label{diagram}
\end{center}
\end{figure}

\begin{figure}
\begin{center}
\includegraphics[scale=0.22]{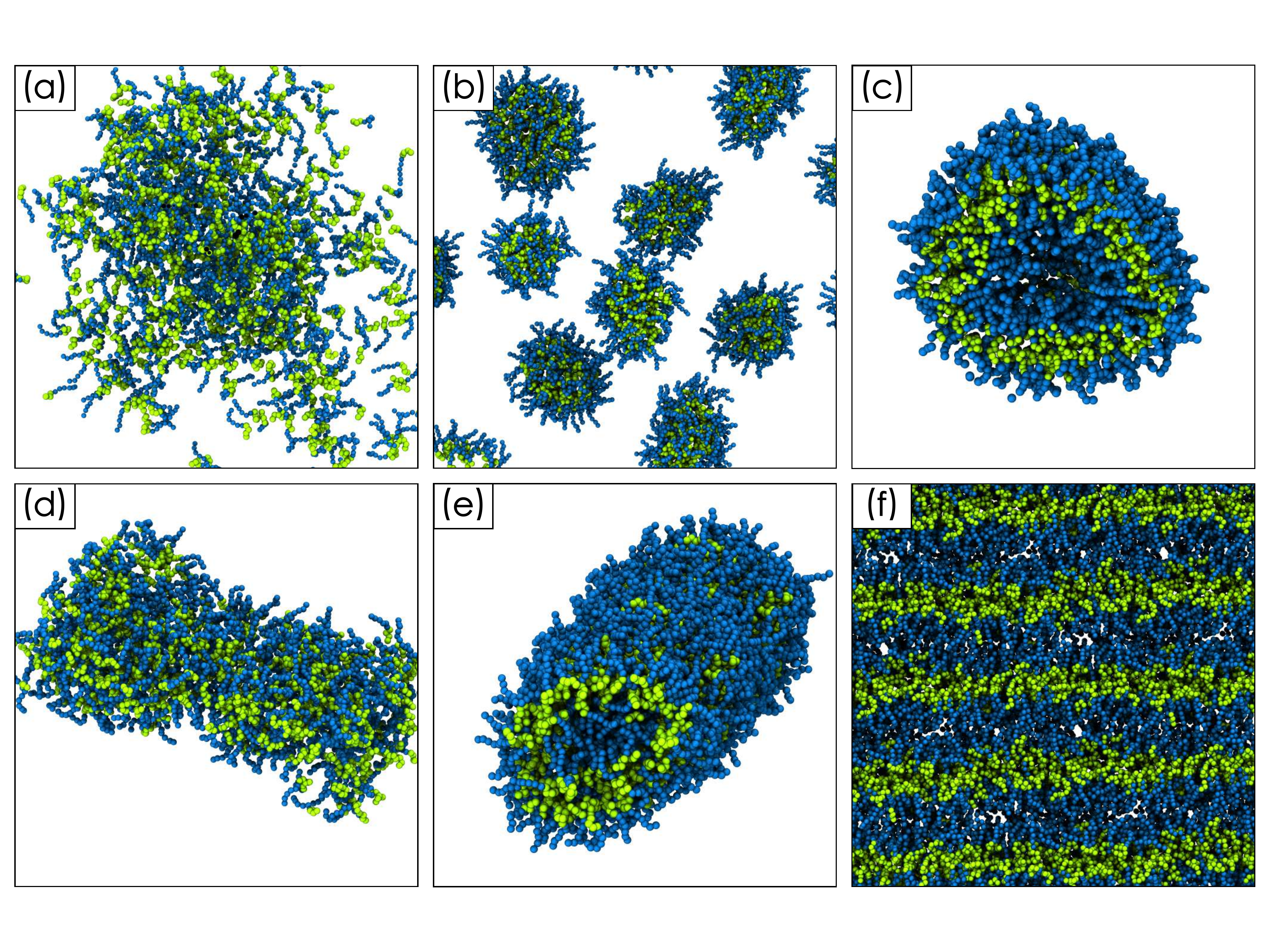}
\caption{Representative snapshots of the morphologies obtained from the self-assembly of PEO$_{6}$-\textit{b}-PBMA$_{4}$ in water and THF: (a) clusters, (b) kinetically-trapped spheres (see text for details), (c) spherical vesicles, (d) dispersed sheets or disk-like aggregates, (e) rod-like vesicles and (f) lamellae. Green and blue beads correspond to PBMA and PEO blocks, respectively. THF and water are not shown for clarity.}
\label{morpho}
\end{center}
\end{figure}

In this section, we present the morphologies resulting from the self-assembly of PEO$_{6}$-\textit{b}-PBMA$_{4}$ in THF and water, and include the analysis of the mesophases containing PEO$_{12}$-\textit{b}-PBMA$_{10}$ in the Supplementary Information. The morphological ternary diagram, unveiling the aggregation behavior of PEO$_{6}$-\textit{b}-PBMA$_{4}$ as function of the common/selective solvent composition, is reported in Fig. \ref{diagram}, while some pictorial representations of the corresponding nanostructures are reported in Fig. \ref{morpho}. In total, we have identified 6 different assembled structures: (i) \textit{clusters}, in which the chains are in close contact to each other, but the hydrophobic and hydrophilic cores are not clearly segregated; (ii) \textit{rod-like vesicles}, being anisotropic hollow structures with a hydrophobic wall, hydrophilic internal and external coronas, and some solvent in the inner core; (iii) \textit{vesicles}, being characterized by a hollow structure with a spherical symmetry; (iv) \textit{sheets or disk-like aggregates}, which are dissolved flat bilayers; (v) \textit{kinetically-trapped spheres}, resembling large crew-cut micelles with fully segregated hydrophobic cores; and (vi) \textit{lamellae}, being ordered arrays of bilayers adjacent to each other and presenting solvent in the interstices. The conditions at which these aggregates are formed are discussed in the remaining of this article.

Following any of the horizontal lines in the ternary diagram of Fig. \ref{diagram}, between $\omega_{\textrm{W}}=0$ and $\omega_{\textrm{THF}}=0$, one can appreciate the effect of the solvent quality on the formation of the equilibrium morphologies. In particular, at a given BCP concentration, between $\omega_{\textrm{BCP}}=0.05$ and 0.50, we find at least one transition involving ordered and disordered mesophases. At low BCP concentrations, the morphological transitions are dominated by changes in the solvent conditions, whereas at relatively high BCP concentrations this effect becomes less evident. At $\omega_{\textrm{BCP}}=0.50$, the copolymer's chains are unable to form long-range order structures at $f_{\textrm{w}} \le 0.20$. However, at $f_{\textrm{w}} = 0.30$, lamellar phases, which remain substantially unchanged upon further addition of water, are observed. A more intriguing behavior is detected at slightly lower BCP concentrations, precisely at $\omega_{\textrm{BCP}}=0.40$, where dispersed finite-size sheets or disk-like aggregates, forming at $f_{\textrm{w}} = 0.50$, are found in between disordered phases and lamellae. Both disk-like aggregates and lamellae are reported in Fig.~\ref{morpho}(d) and (f), respectively.

By gradually decreasing the BCP's concentration, the complexity and richness of the aggregation behavior significantly increase with more morphologies forming. The iso-concentration line $\omega_{\textrm{BCP}}=0.30$ displays dissolved chains or unimers at $f_{\textrm{w}} \le 0.10$, small quasi-spherical clusters at $0.20 \le f_{\textrm{w}} \le 0.50$, dispersed sheets or disk-like aggregates at $f_{\textrm{w}} = 0.60$, and lamellae at $f_{\textrm{w}} \ge 0.70$. The difference between the unimer and disordered state is based solely on the composition: while in the unimer state chains can be considered as being dissolved in a continuous liquid phase of solvents, the disordered state is characterized by a dense phase where the BCP can be regarded as the continuous phase. The clusters of Fig.~\ref{morpho}(a), forming at weak segregation conditions for $0.20 \le f_{\textrm{w}} \le 0.50$, do not present fully-separated hydrophobic and hydrophilic domains, and thus cannot be classified as micelles. A similar set of morphologies are also obtained at $\omega_{\textrm{BCP}}=0.20$, where the stability region of dispersed sheets or disk-like aggregates is significantly larger as compared to $\omega_{\textrm{BCP}}=0.30$. The impact of the solvent composition on the self-assembly of the BCP chains was found to be more pronounced at very low copolymer concentrations. In particular, at $\omega_{\textrm{BCP}}=0.05$, chains are completely dispersed in the solvent for $f_{\textrm{w}} \le 0.40$. Under these conditions, the dissolving effect of THF, which is a good solvent for both PEO and PBMA blocks, dominates over the block-selective interactions that are established by water. However, for $f_{\textrm{w}} > 0.40$, the amount of water is sufficient to promote segregation between the hydrophilic and hydrophobic blocks as a consequence of its repulsive interactions with the bulky PBMA segments. It is very remarkable to notice that the water content at $f_{\textrm{w}} = 0.50$, corresponding to 0.48 wt$\%$, lies within the range of the typical experimental values of the so-called \emph{critical water concentration} (CWC), defined as the minimum water concentration that produces micellization in the presence of a common solvent~\cite{mai2012}. The CWC is analogue to the CMC for aqueous solutions of small surfactants and depends on the properties of both BCP and common solvent~\cite{sarkar2013}. At $0.50 \le f_{\textrm{w}} \le 0.70$, we observe a dispersion of quasi-spherical clusters that, by further addition of water, transform into rod-like vesicles at $f_{\textrm{w}} = 0.80$. These vesicles, reported in Fig.~\ref{morpho}(e), undergo a transition to spherical vesicles at $f_{\textrm{w}} = 0.90$ and eventually evolve into spheres upon further increasing of the water/THF ratio and up to $f_{\textrm{w}} = 1$. The aforementioned morphologies have been experimentally observed on a family of PEO-\emph{b}-PBMA copolymers, whose aggregation behavior in water/THF mixtures was shown to be determined by the proportion of the PEO and PBMA blocks in the chains and the non-selective (common) cosolvent~\cite{mckenzie2013}.

\begin{figure}
\begin{center}
\includegraphics[scale=0.33]{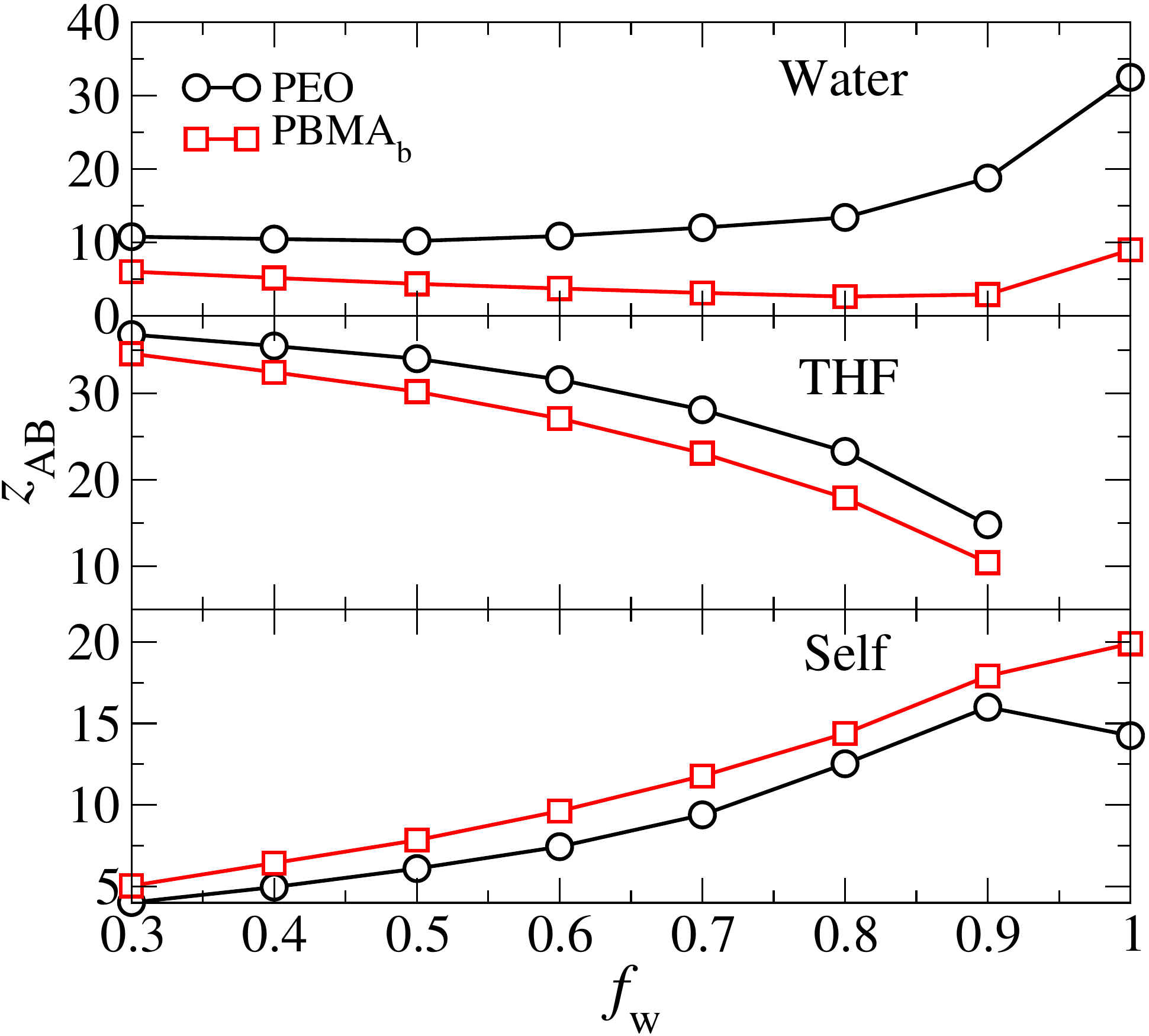}
\caption{Average number of water (top frame), THF (middle frame), and PEO and PBMA$_{\textrm{b}}$ (bottom frame) beads in a sphere of radius $r_{\textrm{c}}=1.2$ nm centred on a PEO (black circles) or PBMA$_{\textrm{b}}$ (red squares) site as a function of $f_\textrm{W}$ at $\omega_{\textrm{BCP}}=0.10$. The solid lines are guides for the eye.}
\label{z1s}
\end{center}
\end{figure}

To gain a clearer understanding on how the relative content of co-solvents influences our system's polymorphism, we now focus on the iso-concentration line $\omega_{\textrm{BCP}}=0.10$. It is well-known that the self-assembly of BCPs is mainly governed by the balance of three free-energy contributions determined by (\textit{i}) the chains' conformation within the aggregate, (\textit{ii}) the interface tension between the hydrophobic core and selective solvent, and (\textit{iii}) the repulsive interactions between the hydrophilic groups in the aggregate's corona~\cite{mai2012}. All these properties can be altered by modifying the ratio between the selective and non-selective solvent. It is therefore crucial to understand how the local environment of both PEO and PBMA blocks changes with $f_\textrm{W}$. To this end, we analyzed the spatial correlations established between these two blocks and the co-solvents. More specifically, given a bead of type A, the average number of beads of type B, $z_{\textrm{AB}}$, that can be found within a spherical shell of radius $r_{\textrm{c}}=1.2$ nm (corresponding to the cut-off distance of the pair interactions) centred on particle A, is obtained from the spatial integration of the pair-correlation function $g_{\textrm{AB}}(r)$ and reads

\begin{equation}
\label{corre}
 z_{\textrm{AB}} = \left< \rho_{\textrm{B}} \right> \int_{0}^{r_{\textrm{c}}}{4\pi r^{2} g_{\textrm{AB}}(r) dr},
\end{equation}

\noindent where $\left< \rho_{\textrm{B}} \right> $ is the number density of beads of type B averaged over all spheres around particles A. In Fig.~\ref{z1s}, we report $z_{\textrm{AB}}$ for the PEO and PBMA$_{\textrm{b}}$ beads  as obtained at $\omega_{\textrm{BCP}}=0.10$ and for a range of common/selective solvent ratios producing aggregates, that is $0.30 \le f_{\textrm{w}} \le 1.0$.

By gradually increasing $f_{\textrm{w}}$, the local distribution of the co-solvents around the two BCP's blocks changes. On the one hand, $z_{\textrm{PEO-THF}}$ and $z_{\textrm{PBMA}_\textrm{b}\textrm{-THF}}$ decrease by approximately 60\% and 70\%, respectively, upon an increase of $f_{\textrm{w}}$ from 0.3 to 0.9. Such a comparable decrease is due to the similar affinity of THF with both BCP blocks. On the other hand, $z_{\textrm{PEO-Water}}$ and $z_{\textrm{PBMA}_\textrm{b}\textrm{-Water}}$ present a rather slight variation with $f_{\textrm{w}}$. In particular, the number of water molecules around the PEO beads slowly increases as a consequence of the removal of THF. Significant changes are detected at $f_{\textrm{w}}>0.8$ where $z_{\textrm{PEO-Water}}$ increases from 15 up to 32 at $f_{\textrm{w}}=1.0$. In contrast, $z_{\textrm{PBMA}_\textrm{b}\textrm{-Water}}$ remains effectively unchanged and exhibits a very small decrease up to $f_{\textrm{w}}=0.9$. This indicates the tendency of the PBMA blocks to avoid the unfavourable contacts with water. Therefore in order to keep $z_{\textrm{PBMA}_\textrm{b}\textrm{-Water}}$ at a minimum value upon the additon of water, the BCP chains start to aggregate into the increasingly packed structures as suggested by the increase in the local number of like sites (bottom frame in Fig.~\ref{z1s}). This mechanism results in the morphological transitions reported in Fig.~\ref{diagram}, which follow a cluster$\rightarrow$sheet$\rightarrow$rod-like vesicle$\rightarrow$spherical vesicle sequence. From previous studies, it has been demonstrated that the aggregate-solvent interfacial area always decreases when BCP assemblies change from spherical clusters to rods to vesicles~\cite{sun2005,wang2017}, suggesting that the morphological transformations arise in part as a thermodynamic response of the system to reduce the interfacial energy component of the total free energy.

\begin{figure}
\begin{center}
\includegraphics[scale=0.33]{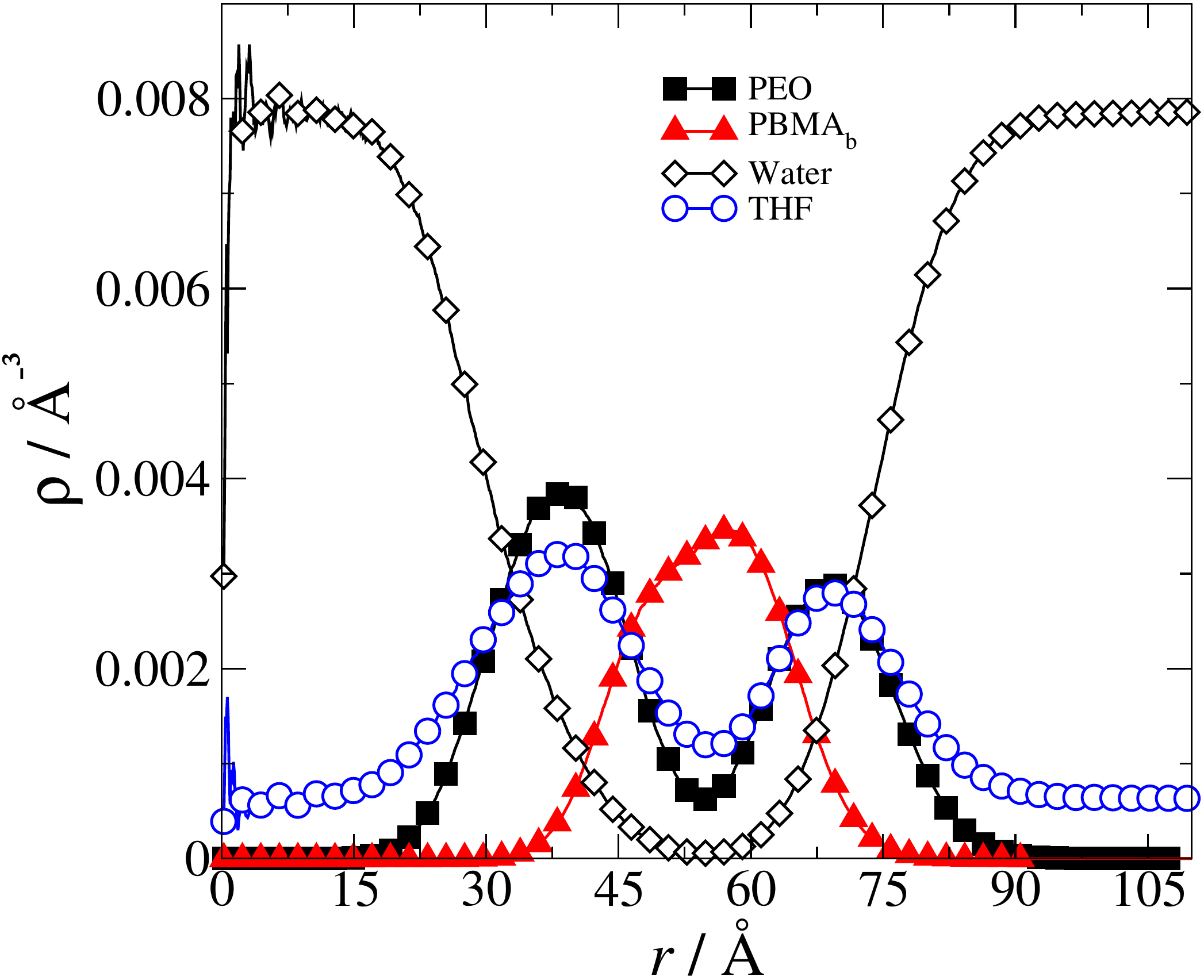}
\caption{Local number density profile of PEO, PBMA$_{\textrm{b}}$, THF and water beads as a function of the radial distance, $r$, from the center of mass of a vesicle at $\omega_{\textrm{BCP}}=0.10$ and $f_{\textrm{w}} = 0.90$.}
\label{rhov}
\end{center}
\end{figure}

It should also be mentioned that while water and THF are macroscopically miscible in the whole range of compositions at $T=300$ K~\cite{smith2016}, the presence of the BCP's aggregates seems to promote a small-scale segregation resulting in the redistribution of THF around the assemblies. This can be appreciated in Fig.~\ref{rhov}, where we report the density profiles of water, THF, PEO and PBMA$_\textrm{b}$ beads as a function of the radial distance from the center of mass of a vesicle formed at $\omega_{\textrm{BCP}}=0.10$ and $f_{\textrm{w}} = 0.90$. From these profiles one can observe that the aggregate is a hollow sphere with a well-defined hydrophobic wall and hydrophilic internal and external coronas, and entrapping water and THF in the interior (see Fig.~\ref{morpho}(c)). The THF density distribution follows very closely that of the PEO block, suggesting a strong spatial correlation between these two type of beads. Consequently, the relative water-to-THF content decreases from 9, in the center of the vesicle, to 0.4 at $r \approx 40$ $\textup{\AA}$, where the inner hydrophilic wall surrounding the internal solventsand indicated by the first peak of $\rho_{\textrm{PEO}}$ is located. The slightly asymmetric profile of $\rho_{\textrm{PEO}}$ is most likely due to the shape fluctuations of the vesicle's cross sectional area, here assumed  circular to estimate the density distribution profiles, but actually ellipsoidal over a significant simulation time window (additional details in the Supporting Information).

Special attention deserve the systems at $f_{\textrm{w}} = 1.0$ and $\omega_{\textrm{BCP}} \le 0.10$. These systems, which are binary mixtures of water and BCP, do not form the spherical vesicles observed at $\omega_{\textrm{THF}}=0.10$, but rather a nearly mono-dispersed population of spherical aggregates, resembling large micelles. According to Johnson and Prud'homme, \textit{micelles} are dynamic systems characterized by fast unimer exchange rates, while \textit{nanoparticles} are kinetically trapped systems, where the unimer exchange rate is extremely slow~\cite{johnson_2003}. Although no inter-aggregate chain exchange has been detected over a significant simulation time, spanning several microseconds, we are aware that in selective solvents the unimer exchange rate can vary between milliseconds and minutes~\cite{nicolai2010}. Consequently, we are not in the position to unambiguously affirm whether the spherical aggregates formed at $f_{\textrm{w}} = 1.0$ are frozen or dynamic objects. Because very slow exchange rates can still spark morphological transitions at very long time scales~\cite{cerritelli_2005}, the relative stability of the spherical aggregates found at $f_{\textrm{w}} = 1.0$ with respect to the (expected) vesicular structures was also assessed by using as starting configuration the vesicular system obtained at $f_{\textrm{w}} = 0.9$. To this end, all the THF beads in this configuration were replaced by water beads and an extra 5 $\mu$s MD trajectory produced. Over this simulation time, no vesicle-to-sphere transition occurred, confirming the key role played by the common solvent in determining the aggregation kinetics. Similarly, by simulating the spheres obtained at $f_{\textrm{w}} = 1.0$ and adding the amount of THF to get $f_{\textrm{w}} = 0.9$, we observed the direct transition to the vesicular structure after the first microsecond. We conclude that the dispersed spheres at $f_{\textrm{w}} = 1.0$, although being highly stable (for about 6 $\mu$s) and monodisperse (with a mean aggregation number $N_{\textrm{agg}}=143$ $\pm11$) might be kinetically-frozen morphologies as we indicate in the ternary diagram of Fig. \ref{diagram}.

\begin{figure}
\begin{center}
\includegraphics[scale=0.22]{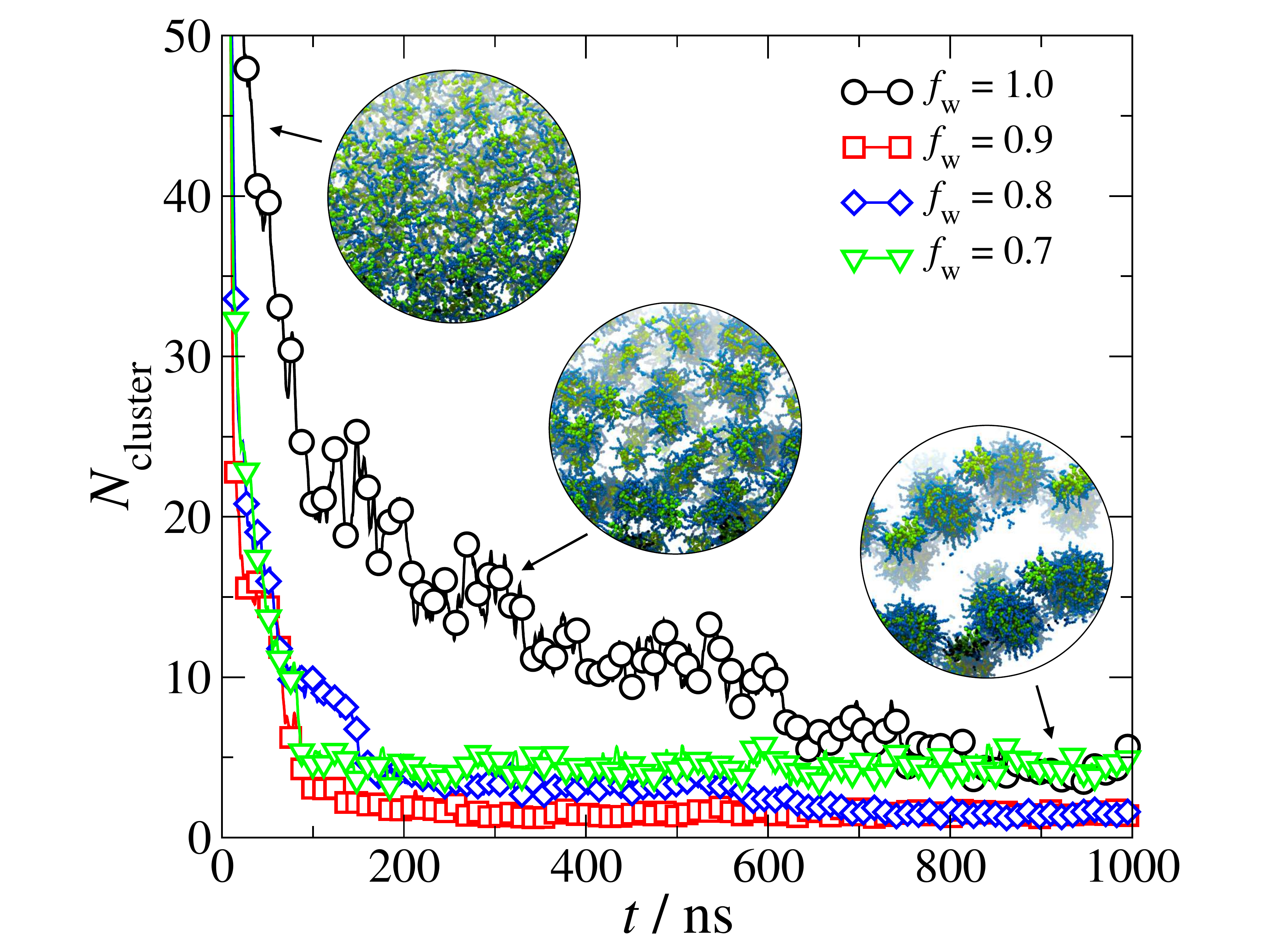}
\caption{Time dependence of the number of clusters forming in water/THF/BCP mixtures at $\omega_{\textrm{BCP}}=0.10$ and $0.70 \le f_{\textrm{w}} \le 1.0$. Chains are dissolved at time $t=0$ ns. The three insets are intermediate configurations of the system with $f_{\textrm{w}}=1.0$.}
\label{CLUSTER}
\end{center}
\end{figure}

To better understand the effect of the non-selective solvent on the self-assembly process, we now focus our attention on the aggregation kinetics by tracking the number of clusters, $N_{\textrm{cluster}}$, as a function of time at varying $f_{\textrm{w}}$. In this work, two chains are considered to be part of the same cluster, if the distance between any of their beads is lower than 1.2 nm. In Fig.~\ref{CLUSTER}, we present the time dependence of $N_{\textrm{cluster}}$ in systems with solvent ratio $0.70 \le f_{\textrm{w}} \le 1.0$. The initial configuration of these systems consists of randomly dissolved unimers that start assembling into small spherical clusters during the first hundreds of nanoseconds. Over larger time scales, such small aggregates merge into larger-sized assemblies and then stabilize into the final morphologies. It is interesting to note that the evolution from the initially dissolved unimers to the equilibrium segregated state occurs at different rates depending on the solvent quality. It is during the first 500 ns that the effect of the presence of THF is more evident. The fast decay of $N_{\textrm{cluster}}$ for $f_{\textrm{w}}=[0.7, 0.9]$ suggests that the BCP chains can more easily diffuse in solution and relax into the self-assembled aggregates as compared to the system without THF. When only water is present, the strong segregation conditions and poor solvent quality for the PBMA blocks impose a kinetic frustration of the components into spherical assemblies and the formation of  vesicular nanostructures is suppressed, at least in the time-scale of our simulations. 

\begin{figure}
\begin{center}
\includegraphics[scale=0.22]{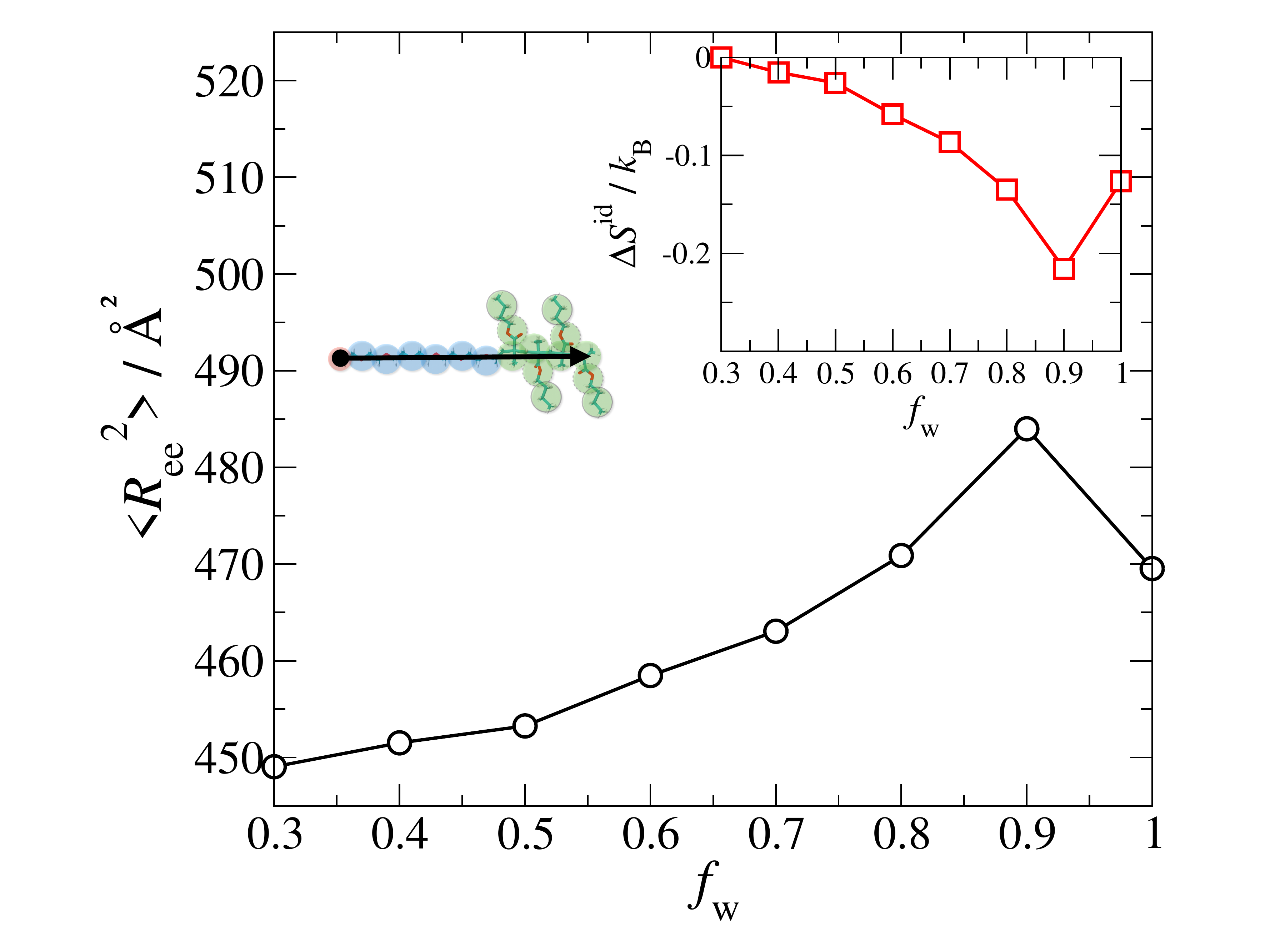}
\caption{Mean-square end-to-end distance of BCP chains as a function of $f_{\textrm{w}}$ at $\omega_{\textrm{BCP}}=0.10$. The corresponding changes in the ideal chain entropy per chain, $\Delta S^{\textrm{id}}$, with respect to the unimer state are reported in the inset. The solid lines are a guide for the eye.}
\label{ends}
\end{center}
\end{figure}

This kinetic frustration might strongly be related to the impact of the solvent quality on the BCP chains' conformation, which ultimately regulates their ability to diffuse from and to the aggregates. It is also known that the morphological transitions in BCP aggregates are influenced by the entropic penalty associated to the reduction in the available conformations due to confinement of the chains within the assemblies~\cite{mai2012}. To determine how the chain conformation changes as a function of the common/selective solvent ratio, we estimated the BCP mean-square end-to-end distance, which reads

\begin{equation}
\label{eq_end}
\left< R_{\textrm{ee}}^{2} \right> =  \left<   \left(   \sum_{i}^{n} \vec{r}_{i} \right) \cdot \left(   \sum_{j}^{n}  \vec{r}_{j} \right) \right> 
\end{equation}

\noindent where $\vec{r}_{i}$ is the bond vector $i$ in the chain of size $N_{\textrm{b}}=n+1$ sites. We compute $ \left< R_{\textrm{ee}}^{2} \right> $ for the whole chain by a summation of the bond vectors over the PEO and PBMA$_{\textrm{b}}$ segments. The resulting dependence of $ \left< R_{\textrm{ee}}^{2} \right> $ on $f_{\textrm{w}}$ is reported in Fig.~\ref{ends}. As a general tendency, $ \left< R_{\textrm{ee}}^{2} \right> $ increases up to $f_{\textrm{w}}=0.9$, indicating that the individual chains gradually stretch when the BCP assemblies undergo morphological transitions in the cluster$\rightarrow$sheet$\rightarrow$rod-like vesicle$\rightarrow$spherical vesicle sequence. However, an inversion is observed at $f_{\textrm{w}} = 1.0$, where kinetically-trapped spheres are formed. These structural changes are in agreement with experimental observations~\cite{zhang1995} and simulation results~\cite{wang2017}. 

By further analyzing the structure of the chains, we note that the ratio between $ \left< R_{\textrm{ee}}^{2} \right> $ and the mean-square radius of gyration, $\left< R_{\textrm{g}}^{2}\right> =1/N_{\textrm{b}} \sum_{i=1}^{N_{\textrm{b}}} \sum_{j=1}^{N_{\textrm{b}}} \left< \left( \vec{R}_{i} - \vec{R}_{j} \right)^{2} \right>$, with $\vec{R}_{i}$ the position vector of site $i$, is in all the cases $ \left< R_{\textrm{ee}}^{2} \right> / \left< R_{\textrm{g}}^{2}\right> \approx 6$
 (see Supporting Information), which agrees with the Debye result of ideal linear chains~\cite{rubinstein2003}. The quasi-ideal conformations adopted by the chains are not specially surprising, because the interactions between BCP sites are highly screened in both the dissolved and aggregated states. Under these considerations, the change in the ideal conformational entropy per BCP chain within the assemblies with respect to the unimer state can be assumed to depend only on $ \left< R_{\textrm{ee}}^{2} \right>$ and roughly approximated to~\cite{rubinstein2003}:
 \begin{equation}
  \Delta S^{\textrm{id}}=3k_{\textrm{B}}\left[ \left< R_{\textrm{ee}}^{2} \right>_{f_{\textrm{w}}=0} - \left< R_{\textrm{ee}}^{2} \right>_{f_{\textrm{w}}} \right]/2N_{\textrm{b}}\left<\sigma \right>^{2},
 \end{equation}
 where $\left<\sigma \right>=4.66 \textup{\AA}$ is the average bead size. The dependence of $\Delta S^{\textrm{id}}$ on $f_{\textrm{w}}$ is reported in the inset of Fig.~\ref{ends}. The balance between this entropically unfavourable effect (increase in the conformational term of the free energy) and the energetic interactions leads to the observed morphological transitions.

Finally, the distinct BCP size of PEO$_{12}$-\textit{b}-PBMA$_{10}$ leads to the formation of significantly different self-assembled structures as observed in the morphological phase diagram reported in the Supplementary Information. This ternary system produced just few common morphologies observed in the THF/water solutions of PEO$_{6}$-\textit{b}-PBMA$_{4}$. These include lamellar phases, which were observed at $\omega_{\text{BCP}} = 0.40$ and solvent ratios $0.10 \leq f_{\text{w}}\leq 0.30$, and kinetically-stabilized spheres, found at the BCP concentrations of $\omega_{\text{BCP}} = 0.10$ and $\omega_{\text{BCP}} = 0.20$, only in the presence of water. The curve of the CWC, which defines the boundary between the disordered and aggregated states, closely follows that of the PEO$_{6}$-\textit{b}-PBMA$_{4}$ diagram. Other interesting meshophases such as rod-like micelles and distorted lamellae appear at intermediate values of $f_{\textrm{w}}$ and BCP contents. More importantly, we detected a large region where bicontinuous structures are formed. In a bicontinuous phase, a twisted network of the hydrophobic blocks intertwines with that of the hydrated hydrophilic segments. Sommerdijk and coworkers recently reported the formation of bicontinuous structures from the spontaneous self-assembly of a family of PEO-\textit{b}-PBMA diblock copolymers in water/THF mixtures~\cite{mckenzie2013}. They found that the PEO content was critical in the stabilization of such morphology as BCPs with PEO contents of $M_{\textrm{PEO}}/M_{\textrm{BCP}}=$0.14 and 0.17 allowed for its formation. In contrast, BCPs with PEO contents of 0.40 did not exhibit this mesophase. In our case, PEO$_{12}$-\textit{b}-PBMA$_{10}$, with a PEO content of 0.27, produces bicontinuous structures whereas PEO$_{6}$-\textit{b}-PBMA$_{4}$, with a relatively larger PEO content (0.31) does not present such a phase. Thus our simulation results also point to the importance of the molecular weight and hydrophilic/hydrophobic balance on the self-assembling behavior.

\section{Conclusions}
In summary, we have performed extensive MD simulations to investigate the self-assembly of PEO$_{6}$-\textit{b}-PBMA$_{4}$ and PEO$_{12}$-\textit{b}-PBMA$_{10}$ copolymers in binary mixtures of water (selective solvent) and THF (common solvent). The dependence of the aggregate morphology on the molecular weight and hydrophilic/hydrophobic ratio of the copolymer chains has been evidenced, with the general observation that finite-size aggregates including spherical clusters, dispersed sheets or disk-like aggregates, anisotropic rod-like vesicles and spherical vesicles are predominantly formed from PEO$_{6}$-\textit{b}-PBMA$_{4}$, which is characterized by a lower hydrophilic content of $M_{\textrm{PEO}}/M_{\textrm{BCP}}=$0.27 with respect to the 0.31 of PEO$_{12}$-\textit{b}-PBMA$_{10}$. The latter exhibits mainly distorted lamellae, rod-like micelles, spheres and more importantly, a large portion of the non-conventional and technologically attractive bicontinuous structures. The majority of these mesophases have been experimentally obtained from the self-assembly of PEO-\textit{b}-PBMA BCPs in mixtures of water and THF~\cite{mckenzie2013}. Furthermore, we have demonstrated that the equilibrium morphology of the self-assembled aggregates is not solely determined by the BCP chain size and concentration. Our simulation results unambiguously indicate that for a fixed BCP content and size, tuning the solvent conditions, in particular the selective/common solvent ratio, is sufficient to induce reversible morphological transitions of the aggregates. This is a consequence of the changes in the spatial correlations between the solvent molecules and each of the copolymer blocks as the mixture composition is modified. Our results also confirm that the morphological transformations promoted by an increase in the amount of selective solvent (water) are accompanied by an entropically unfavourable effect associated to the systematic stretching of the individual chains. Finally, we have seen that the presence of the common solvent strongly influences the self-assembly mechanism by facilitating the aggregation of the chains and thus minimizing the possible formation of kinetically-frozen assemblies.

\section*{Acknowledgements}
The authors acknowledge useful discussions on the experimental methods with Dr Joe Patterson (University of California, Irvine) and Paula Vena (Eindhoven University of Technology). The project leading to these results has received funding from the European Union's Horizon 2020 research and innovation programme under the Marie Sk\l{}odowska-Curie grant agreement No 676045 (MULTIMAT).

\section*{Appendix A. Supporting Information}
Detailed information on the employed force-field parameters, simulation systems, pair-correlation functions, vesicle shape fluctuations and PEO$_{12}$-\textit{b}-PBMA$_{10}$/water/THF phase diagram can be found in the online version, at http://

\end{sloppypar}

\section*{References}

\bibliography{Manuscript}



\section*{Supplementary material}

# Interaction parameters

Parameters for Lennard-Jones (LJ) pair interactions between all types of beads are reported in Tables 1 and 2. Harmonic bond and angle parameters are listed in Table 3. A detailed description of the models can be consulted in Ref.[1]

Table 1: LJ potential parameters for pair interactions of THF and standard MARTINI water. Taken from Ref.[1]

| Pair | $\epsilon$ (kJ mol$^{-1}$) | $\sigma$ (nm) |
|---|---|---|
| THF−THF | 4.000 | 0.480 |
| THF−Water | 4.120 | 0.470 |

Table 2: Force-field parameters for LJ interactions between CG sites comprising the copolymer chains and THF.

| Pair | $\epsilon$ (kJ mol$^{-1}$) | $\sigma$ (nm) | Pair | $\epsilon$ (kJ mol$^{-1}$) | $\sigma$ (nm) |
|---|---|---|---|---|---|
| THF−THF | 4.000 | 0.480 | PEO$_e$−PEO | 3.375 | 0.430 |
| PEO$_e$−PEO$_e$ | 3.750 | 0.430 | PEO$_e$−PBMA$_b$ | 2.025 | 0.430 |
| PEO−PEO | 3.375 | 0.430 | PEO$_e$−PBMA$_e$ | 2.025 | 0.430 |
| PBMA$_b$−PBMA$_b$ | 2.625 | 0.430 | PEO$_e$−PBMA$_c$ | 4.500 | 0.470 |
| PBMA$_e$−PBMA$_e$ | 2.625 | 0.430 | | | |
| PBMA$_c$−PBMA$_c$ | 4.000 | 0.470 | PEO−PBMA$_b$ | 2.025 | 0.430 |
| | | | PEO−PBMA$_e$ | 2.025 | 0.430 |
| THF−PEO$_e$ | 4.500 | 0.470 | PEO−PBMA$_c$ | 4.500 | 0.470 |
| THF−PEO | 4.500 | 0.470 | | | |
| THF−PBMA$_b$ | 2.700 | 0.470 | PBMA$_b$−PBMA$_e$ | 2.625 | 0.430 |
| THF−PBMA$_e$ | 2.700 | 0.470 | PBMA$_b$−PBMA$_c$ | 2.700 | 0.470 |
| THF−PBMA$_c$ | 4.500 | 0.470 | | | |
| | | | PBMA$_e$−PBMA$_c$ | 2.700 | 0.470 |

Table 3: Force-field parameters for the intramolecular interactions of PEO-$b$-PBMA chains.

| Bonded Interaction | $l_0$ (nm) / $\theta_0$ (deg) | $K_l$ (kJ mol$^{-1}$ nm$^{-2}$) / $K_\theta$ (kJ mol$^{-1}$) |
|---|---|---|
| Bonds | | |
| PEO$_e$-PEO | 0.299 | 12500 |
| PEO-PEO | 0.348 | 8500 |
| PEO-PBMA$_b$ | 0.335 | 16300 |
| PBMA$_b$-PBMA$_b$ | 0.289 | 21100 |
| PBMA$_b$-PBMA$_c$ | 0.282 | 17000 |
| PBMA$_e$-PBMA$_c$ | 0.383 | 5000 |
| | | |
| Angles | | |
| PEO$_e$-PEO-PEO | 180 | 37 |
| PEO-PEO-PEO | 170 | 40 |
| PEO-PEO-PBMA$_b$ | 180 | 30 |
| PEO-PBMA$_b$-PBMA$_b$ | 139 | 15 |
| PBMA$_b$-PBMA$_b$-PBMA$_b$ | 175 | 13 |
| PBMA$_b$-PBMA$_c$-PBMA$_e$ | 144 | 67 |

# Simulation systems

Table 4: Simulated systems indicating the BCP content, solvent ratio ($f_w$), water content and THF content. The respective number of solvent beads in each case are also reported in brackets for $PEO_6$-$b$-$PBMA_4$ and $PEO_{12}$-$b$-$PBMA_{10}$, respectively.

| BCP Content (wt%) | $f_w$ | Water Content (wt%) | THF Content (wt%) |
|---|---|---|---|
| 5.0 | 0.1 | 9.5 (16466) | 85.5 (148200) |
|  | 0.2 | 19.0 (32933) | 76.0 (131733) |
|  | 0.3 | 28.5 (49399) | 66.5 (115267) |
|  | 0.4 | 38.0 (65866) | 57.0 (98800) |
|  | 0.5 | 47.5 (82333) | 47.5 (82333) |
|  | 0.6 | 57.0 (98800) | 38.0 (65866) |
|  | 0.7 | 66.5 (115267) | 28.5 (43399) |
|  | 0.8 | 76.0 (131733) | 19.0 (32933) |
|  | 0.9 | 85.5 (148200) | 9.5 (16466) |
|  | 1.0 | 95.0 (164666) | 0.0 (0) |
| 10.0 | 0.1 | 9.0 (7800) / (18262) | 81.0 (70200) / (164363) |
|  | 0.2 | 18.0 (15600) / (36525) | 72.0 (62400) / (146100) |
|  | 0.3 | 27.0 (23400) / (54787) | 63.0 (54600) / (127838) |
|  | 0.4 | 36.0 (31200) / (73050) | 54.0 (46800) / (109575) |
|  | 0.5 | 45.0 (39000) / (91312) | 45.0 (39000) / (91313) |
|  | 0.6 | 54.0 (46800) / (109575) | 36.0 (31200) / (73050) |
|  | 0.7 | 63.0 (54600) / (127838) | 27.0 (23400) / (54787) |
|  | 0.8 | 72.0(62400) / (146100) | 18.0 (15600) / (36525) |
|  | 0.9 | 81.0 (70200) / (164363) | 9.0 (7800) / (18262) |
|  | 1.0 | 90.0 (78000) / (182625) | 0.0 (0) / (0) |
| 20.0 | 0.1 | 8.0 (3466) / (8116) | 72.0 (31200) / (73050) |
|  | 0.2 | 16.0 (6933) / (16233) | 64.0 (27733) / (64933) |
|  | 0.3 | 24.0 (10399) / (24350) | 56.0 (24267) / (56816) |
|  | 0.4 | 32.0 (13866) / (32466) | 48.0 (20800) / (48700) |
|  | 0.5 | 40.0 (17333) / (40583) | 40.0 (17333) / (40583) |
|  | 0.6 | 48.0 (20800) / (48700) | 32.0 (13866) / (32466) |
|  | 0.7 | 56.0 (24267) / (56816) | 24.0 (10399) / (24350) |
|  | 0.8 | 64.0 (27733) / (64933) | 16.0 (6933) / (16233) |
|  | 0.9 | 72.0 (31200) / (73050) | 8.0 (3466) / (8116) |
|  | 1.0 | 80.0 (34666) / (81166) | 0.0 (0) / (0) |
| 30.0 | 0.1 | 7.0 (2022) / (4734) | 63.0 (18200) / (42613) |
|  | 0.2 | 14.0 (4044) / (9469) | 56.0 (16178) / (37878) |
|  | 0.3 | 21.0 (6066) / (14204) | 49.0 (14156) / (33143) |
|  | 0.4 | 28.0 (8088) / (18938) | 42.0 (12134) / (28409) |
|  | 0.5 | 35.0 (10111) / (23673) | 35.0 (10111) / (23674) |
|  | 0.6 | 42.0 (12134) / (28409) | 28.0 (8088) / (18938) |
|  | 0.7 | 49.0 (14156) / (33143) | 21.0 (6066) / (14204) |
|  | 0.8 | 56.0 (16178) / (37878) | 14.0 (4044) / (9469) |
|  | 0.9 | 63.0 (18200) / (42613) | 7.0 (2022) / (4734) |
|  | 1.0 | 70.0 (20222) / (47347) | 0.0 (0) / (0) |
| 40.0 | 0.1 | 6.0 (1300) / (3043) | 54.0 (11700) / (27394) |
|  | 0.2 | 12.0 (2600) / (6087) | 48.0 (10400) / (24350) |
|  | 0.3 | 18.0 (3900) / (9131) | 42.0 (9100) / (21306) |
|  | 0.4 | 24.0 (5200) / (12175) | 36.0 (7800) / (18262) |
|  | 0.5 | 30.0 (6500) / (15218) | 30.0 (6500) / (15219) |
|  | 0.6 | 36.0 (7800) / (18262) | 24.0 (5200) / (12175) |
|  | 0.7 | 42.0 (9100) / (21306) | 18.0 (3900) / (9131) |
|  | 0.8 | 48.0 (10400) / (24350) | 12.0 (2600) / (6087) |
|  | 0.9 | 54.0 (11700) / (27394) | 6.0 (1300) / (3043) |
|  | 1.0 | 60.0 (13000) / (30437) | 0.0 (0) / (0) |
| 50.0 | 0.1 | 5.0 (866) | 45.0 (7800) |
|  | 0.2 | 10.0 (1733) | 40.0 (6933) |
|  | 0.3 | 15.0 (2599) | 35.0 (6067) |
|  | 0.4 | 20.0 (3466) | 30.0 (5200) |
|  | 0.5 | 25.0 (4333) | 25.0 (4333) |
|  | 0.6 | 30.0 (5200) | 20.0 (3466) |
|  | 0.7 | 35.0 (6067) | 15.0 (2599) |
|  | 0.8 | 40.0 (6933) | 10.0 (1733) |
|  | 0.9 | 45.0 (7800) | 5.0 (866) |
|  | 1.0 | 50.0 (8666) | (0) |

# Pair-correlation functions and solvent distributions

The pair-correlation function between species A and B is calculated as:

$$g_{\mathrm{AB}}(r) = \frac{1}{\langle \rho_{\mathrm{B}} \rangle} \frac{1}{N_{\mathrm{A}}} \sum_{i \in \mathrm{A}}^{N_{\mathrm{A}}} \sum_{j \in \mathrm{B}}^{N_{\mathrm{B}}} \frac{\delta(r_{ij} - r)}{4\pi r^2}, \qquad (1)$$

where $\langle \rho_{\mathrm{B}} \rangle$ is the number density of particle B averaged over all spheres centred on particle A with radius $r_{\mathrm{sp}} = 3.0$ nm. The resulting curves are reported in Fig. 1.

Two-dimensional probability density maps of THF and water beads around a self-assembled vesicle formed from $\mathrm{PEO}_6$-$b$-$\mathrm{PBMA}_4$ at $\omega_{\mathrm{BCP}} = 0.1$ and $f_{\mathrm{w}} = 0.9$ are presented in Fig. 2.

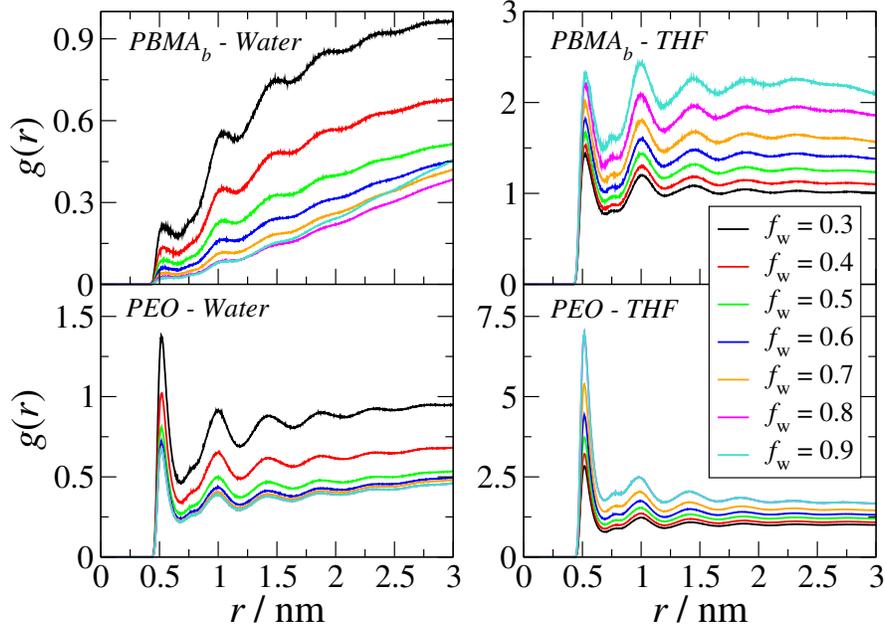

Figure 1: Pair-correlation function between distinct bead types at $\omega_{\mathrm{BCP}} = 0.10$ and distinct $f_{\mathrm{w}}$ values.

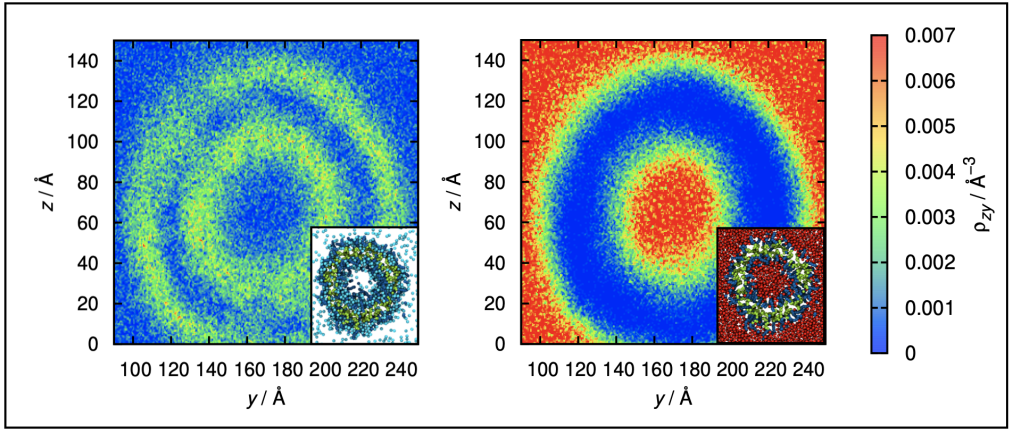

Figure 2: Two-dimensional probability density maps of THF (left) and water (right) around a self-assembled vesicle formed from PEO$_6$-$b$-PBMA$_4$ at $\omega_{\text{BCP}} = 0.1$ and $f_{\text{w}} = 0.9$. The 2D plane intersects the center of mass of the vesicle. The small insets correspond to typical simulation snapshots.

## Vesicle shape fluctuations

An interesting feature observed during the extensive MD simulations of the vesicles is the marked thermal deformations they experience in solution. In order to track these shape fluctuations we compute the asphericity parameter $b$, which measures the deviation from spherical symmetry. This shape descriptor is calculated as:

$$b = \lambda_1 - \frac{1}{2}\left(\lambda_2 + \lambda_3\right), \qquad (2)$$

where $\lambda_1 \geq \lambda_2 \geq \lambda_3$ are the eigenvalues of the gyration tensor $S_{mn} = \frac{1}{N}\sum_{i=1}^{N} r_m^{(i)} r_n^{(i)}$. The first invariant of $S_{mn}$ gives the squared radius of gyration,

$$\text{Tr} S_{mn} = \lambda_1 + \lambda_2 + \lambda_3 = R_{\text{g}}^2. \qquad (3)$$

Another useful quantity to measure deviation from highly symmetric body conformations

is the second invariant shape descriptor, or relative shape anisotropy, defined as

$$\kappa^2 = 1 - 3\frac{\lambda_1\lambda_2 + \lambda_2\lambda_3 + \lambda_3\lambda_1}{(\lambda_1 + \lambda_2 + \lambda_3)^2}. \quad (4)$$

The time dependence of $b$ and $\kappa^2$ are reported in Fig. 3, whereas the corresponding eigenvalues of the gyration tensor are given in Fig. 4.

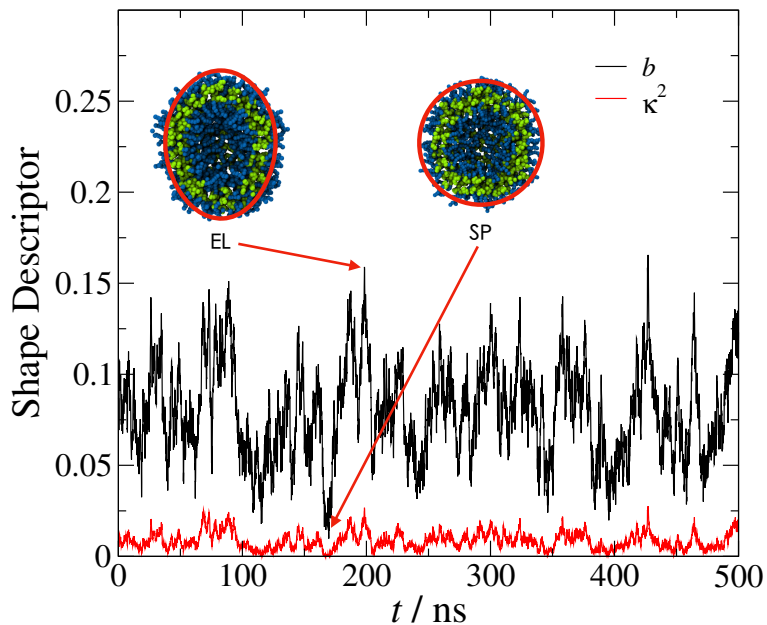

Figure 3: Time variation of the asphericity parameter, $b$, and relative shape anisotropy, $\kappa^2$ of a self-assembled vesicle formed from $PEO_6$-$b$-$PBMA_4$ at $\omega_{BCP} = 0.1$ and $f_w = 0.9$. The snapshots represent instantaneous simulation configurations of the vesicles in the elongated (EL) and (quasi) spherical conformations (SP).

7

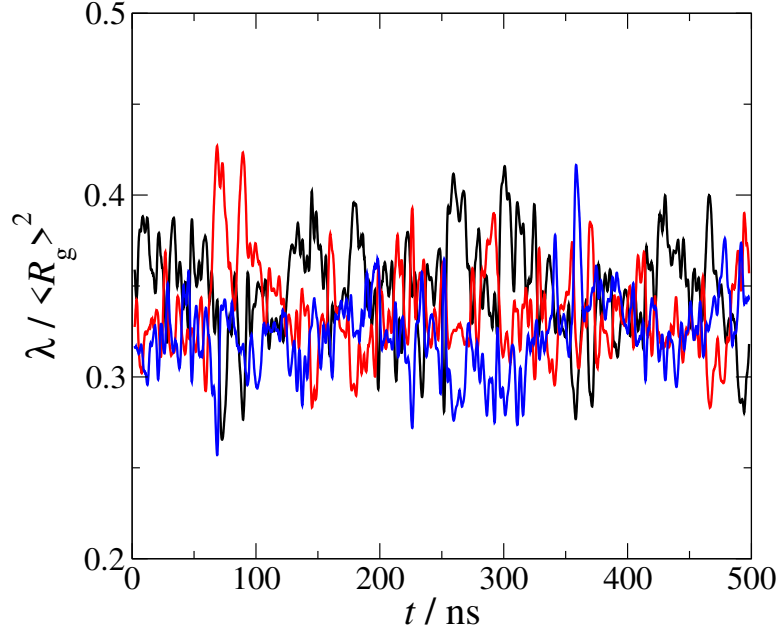

Figure 4: Time variation of the eigenvalues of the gyration tensor ($\lambda_i$) of a self-assembled vesicle formed from PEO$_6$-$b$-PBMA$_4$ at $\omega_{\text{BCP}} = 0.1$ and $f_{\text{w}} = 0.9$.

# Morphological phase diagram of PEO$_{12}$-$b$-PBMA$_{10}$

The morphological phase diagram of PEO$_{12}$-$b$-PBMA$_{10}$ in mixtures of water and THF at $T = 300$ K is reported in Fig. 5. As in the case of PEO$_6$-$b$-PBMA$_4$, lamellae (L) and kinetically-trapped spheres (M) are observed. In addition 4 different morphologies appear, namely (i) *rod-like micelles* (R), corresponding to elongated core-shell aggregates, (ii) *distorted lamella* (DL), which are non-flat bilayers, (iii) *octopus-like aggregates* (O), which are characterized by cylindrical micellar protrusions converging in a small flattened bilayer sheet and (iv) *bicontinuous phases* (B), in which a twisted network of the hydrophobic blocks intertwines with that of the hydrated hydrophilic segments.

8

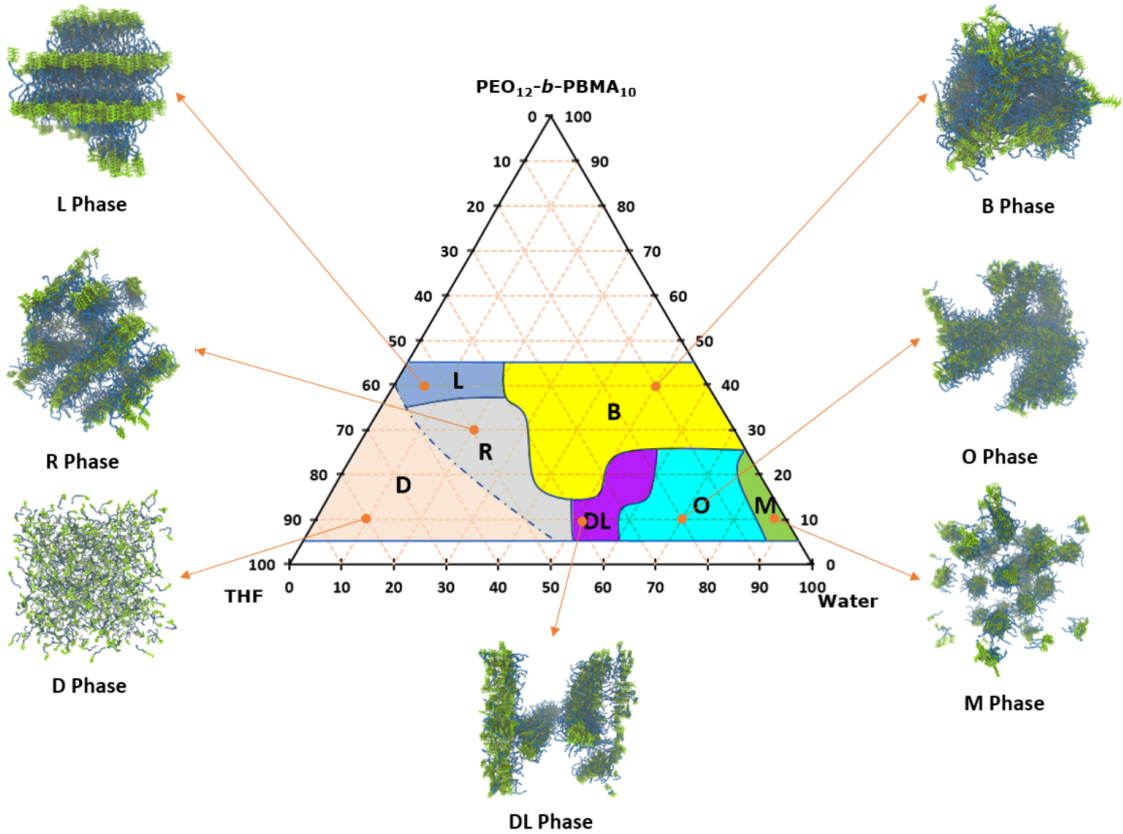

Figure 5: Morphological phase diagram of PEO$_{12}$-$b$-PBMA$_{10}$ in mixtures of water and THF at $T = 300$ K. The shaded regions indicate the ternary mixture compositions in which lamellae (L), rod-like micelles (R), dispersed chains (D), distorted lamellae (DL), kinetically-trapped spheres (M), octopus-like aggregates (O) and bicontinuous phases (B) are obtained. Simulation snapshots of the corresponding morphologies are illustrated with the PEO and PBMA blocks represented with blue and green spheres, respectively.

# BCP chain structure

The ratio between the mean square end-to-end distance and mean square radius of gyration, $\langle R_{\text{ee}}^2 \rangle / \langle R_{\text{g}}^2 \rangle$ of PEO$_6$-$b$-PBMA$_4$ at $\omega_{\text{BCP}} = 0.1$ and different solvent ratios, $f_{\text{w}}$, is reported in Fig. 6.

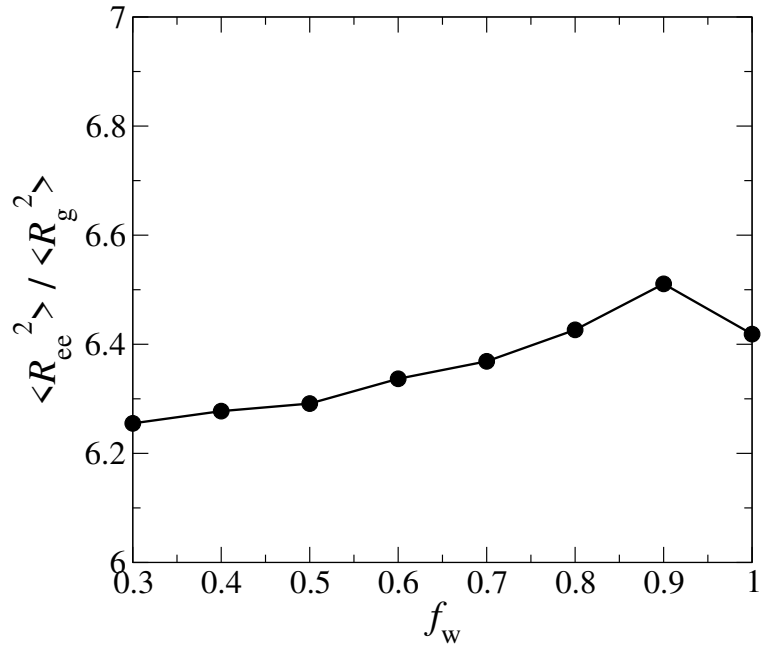

Figure 6: Ratio between the mean square end-to-end distance and mean square radius of gyration, $\langle R_{ee}^2 \rangle / \langle R_g^2 \rangle$ of $PEO_6$-$b$-$PBMA_4$ at $\omega_{BCP} = 0.1$.



\end{document}